\documentstyle[floats,aps,epsf,prl,psfig]{revtex}
\begin{document}
\draft
\twocolumn[\hsize\textwidth\columnwidth\hsize\csname @twocolumnfalse\endcsname
\title{Structure, barriers and relaxation mechanisms of kinks in the
90$^{\circ}$ partial dislocation in silicon}

\author{R. W. Nunes, J. Bennetto, and David Vanderbilt}

\address{Department of Physics and Astronomy, 
Rutgers University, Piscataway, NJ 08855-0849}

\date{\today}
\maketitle

\begin{abstract}

Kink defects in the 90$^{\circ}$ partial dislocation in silicon are
studied using a linear-scaling density-matrix technique. The
asymmetric core reconstruction plays a crucial role, generating at
least four distinct kink species as well as soliton defects. The
energies and migration barriers of these entities are calculated and
compared with experiment. As a result of certain low-energy kinks, a
peculiar alternation of the core reconstruction is predicted.  We find
the solitons to be remarkably mobile even at very low temperature, and
propose that they mediate the kink relaxation dynamics.

\end{abstract}


\vskip2pc]
\narrowtext

The importance of dislocations in semiconductors hardly needs comment.
In addition to being responsible for plastic behavior in general,
dislocations occur commonly at semiconductor interfaces where they can
act as trapping and scattering centers for carriers. In silicon, the
predominant slip system consists of 60$^{\circ}$-edge and screw
dislocations oriented along $\langle 110\rangle$ and lying in a
$\{111\}$ slip plane.  Both are known to dissociate into pairs of
partial dislocations bounding a ribbon of stacking fault
\cite{hirsch}. The resulting 90$^{\circ}$ or 30$^{\circ}$ partial
dislocations are believed to have reconstructed cores, consistent with
the low density of dangling bonds as observed by EPR measurements
\cite{hirsch}. Since the dislocation motion occurs by nucleation and
propagation of kinks along the dislocation line, a detailed
understanding of the atomic-scale structure of the kinks is obviously
of the greatest importance. Unfortunately, experimental approaches
have not proved capable of providing such an understanding.

Until recently, the only theoretical methods capable of treating such
problems were based on classical interatomic potentials. These are of
questionable accuracy, and are generally unable to reproduce effects
of intrinsic quantum-mechanical nature such as bond reconstruction and
Peierls or Jahn-Teller symmetry breaking. For example, while the
Stillinger-Weber\cite{stwb} potential has been used to study the core
reconstruction and kinks of the 30$^{\circ}$ partial\cite{bulatov}, it
fails to reproduce the correct core reconstruction of the 90$^{\circ}$
partial\cite{bigger}. It is thus exciting to find that {\it ab-initio}
methods are approaching the point of addressing some interesting
questions about dislocations. Recent theoretical work has focused on
such issues as the core reconstruction of the 90$^{\circ}$
\cite{bigger,hansen}, and the elastic interaction between dislocations
of a dipole in the shuffle \cite{arias} and glide \cite{hansen}
sets. One first-principles study has even been done on a kink barrier
in the 30$^{\circ}$ partial,\cite{huang} but it assumed a kink
structure previously proposed, and used a relatively small
supercell. Cluster calculations on the 90$^{\circ}$ partial have also
been reported \cite{jones}. However, a comprehensive study of
dislocation kink structure and dynamics would require the use of very
large supercells, for which the application of {\it ab-initio}
techniques is still computationally prohibitive.  Thus, there is a
pressing need for the application of more efficient quantum-mechanics
based methods to study the electronic and structural excitations in
this system.

In this Letter, we use a total-energy tight-binding (TBTE) description
of the electronic and interatomic forces to carry out a detailed
atomistic study of this kind for the 90$^{\circ}$ partial dislocation
in silicon.  The key to making the calculations tractable is our use
of a ``linear-scaling'' or ``${\cal{O}}(N)$'' method of solution of
the Schr\"{o}dinger equation \cite{lnv}, enabling us to treat system
sizes up to $10^3$ atoms easily on a workstation platform.  We verify
that the dislocation core reconstructs with a spontaneous symmetry
breaking, and find that the ``soliton'' defect associated with the
reversal of the core reconstruction is extremely mobile.  We find that
at least four distinct kink structures must be considered (labeled by
the sense of the core reconstruction on either side), and show that
they can be classified as high- or low-energy kinks depending on
whether or not they contain a dangling bond.  Molecular-dynamics
simulations as well as fully-relaxed static calculations are used to
characterize formation energies, migration barriers, and kink-soliton
reaction pathways.  The picture that emerges is one in which the
ground state is free of dangling bonds (even in the presence of
kinks), and in which the solitons mediate the structural excitations
and dynamics.

We use the TBTE parameters of Kwon {\it et al.} \cite{kwon}, with a
real-space cutoff of 6.2\AA\ on the range of the density matrix used
in the ${\cal{O}}(N)$ method \cite{lnv}. We chose to work at a fixed
electron chemical potential 0.4 eV above the valence-band edge, and
thus all ``energies'' reported below are technically values of grand
potential \cite{lnv}.  The numerical minimization of the $\cal{O}(N)$
functional was carried out by the conjugate-gradient algorithm, with
the internal line minimization performed exactly \cite{exact}.
Ground-state structures were computed by allowing all atomic
coordinates to relax fully (forces less than 5 meV/\AA).  The
supercells used will be described below; all energies for defect
(soliton and kink) structures are given with respect to a
corresponding supercell containing defect-free (but reconstructed and
fully relaxed) dislocations.  Barrier energies were calculated by
choosing a reaction coordinate and, for a series of values of this
coordinate, computing the energy with this coordinate fixed and all
others fully relaxed. Molecular dynamics runs were performed with a
Verlet algorithm, the Nose thermostat \cite{nose}, and a time step of
2 fs.

In Fig.~1(a), a top view of the atomic structure of the reconstructed
90$^{\circ}$ partial in its slip plane is shown.  
\begin{figure}
\centerline{\psfig{figure=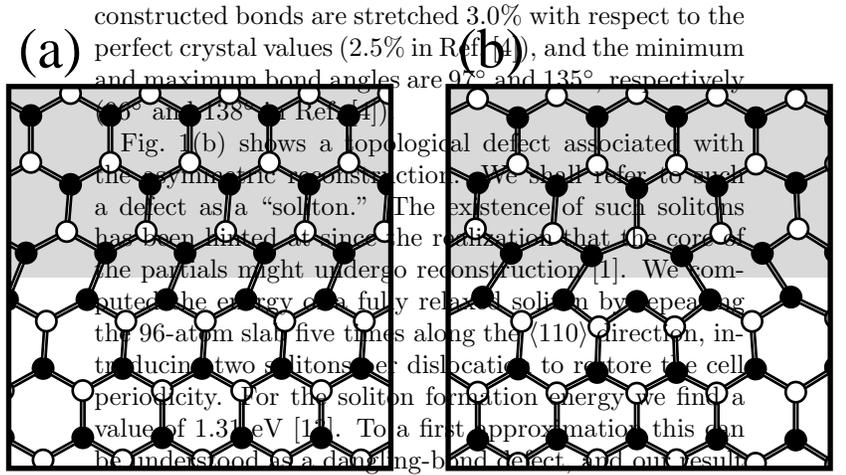,scale=0.7}}
\caption{ 
(a) Top view of the slip plane of a reconstructed 90$^{\circ}$ partial
dislocation. The shaded area indicates the stacking fault. Horizontal
and vertical directions correspond to $[1\overline{1}0]$ and 
$[11\overline{2}]$ directions, respectively. 
(b) Reconstruction defect, or soliton,
where the core reconstruction changes orientation.}
\end{figure}
The shaded area
indicates the stacking fault. The fourfold coordination of the atoms
at the core is restored by bonding across the dislocation line. This
reconstruction has been discussed by other authors
\cite{bigger,hansen} and compared with an alternative
``quasi-fivefold'' bond reconstruction that preserves the mirror
symmetries along the dislocation line. As a test of our TBTE-DM
approach, we computed the energy difference between the two
possibilities.  In this calculation the supercell consisted of a
96-atom slab normal to a $\langle 110\rangle$ direction, containing
two dislocations with opposite burgers vectors separated by a distance
of 13.3 \AA. We find the fourfold reconstruction to be 0.18 eV/\AA\
lower in energy than the fivefold one, in perfect agreement with the
TB calculations in Ref.\cite{hansen}. Our results also compare
favorably with the {\it ab initio} results in Ref.\cite{bigger}, for
which the energy difference between the two reconstructions is 0.23
eV/\AA. In our calculations, the reconstructed bonds are stretched
3.0\% with respect to the perfect crystal values (2.5\% in
Ref.\cite{bigger}), and the minimum and maximum bond angles are
97$^{\circ}$ and 135$^{\circ}$, respectively (96$^{\circ}$ and
138$^{\circ}$ in Ref.\cite{bigger}).

Fig.~1(b) shows a topological defect associated with the asymmetric
reconstruction. We shall refer to such a defect as a ``soliton.'' The
existence of such solitons has been hinted at since the realization
that the core of the partials might undergo
reconstruction\cite{hirsch}. We computed the energy of a fully relaxed
soliton by repeating the 96-atom slab five times along the $\langle
110\rangle$ direction, introducing two solitons per dislocation to
restore the cell periodicity.  For the soliton formation energy we
find a value of 1.31 eV \cite{jonesexplan}.  To a first approximation
this can be understood as a dangling-bond defect, and our result is on
the order of the energy associated with a dangling {\it sp}$^3$
orbital in silicon.

An interesting question is whether a soliton can move easily along the
dislocation. We computed an energy barrier of only 0.04 eV for the
propagation of a soliton between two adjacent equilibrium positions,
as indicated in Fig.~1(b). With such a small barrier, it might be
expected that the solitons would be extremely mobile even at very low
temperatures.  To test this, we performed a molecular dynamics
simulation on a supercell having a soliton-antisoliton pair, initially
separated by 9.6 \AA, on an otherwise defect-free partial dislocation.
Remarkably, at a temperature of only 50 K$^{\circ}$, the solitons were
indeed mobile and recombination of the pair took place after only 1.3
ps. Such highly mobile solitons play an interesting role in the
relaxation of high-energy kinks, as explained below.

A schematic view of the supercell we used for simulating kinks and
soliton-kink complexes is shown in Fig.~2. It contains a total of 864
atoms, corresponding to the 96-atom slab repeated nine times along the
dislocation line ($[1{\overline 1}0]$ direction).  The supercell
vectors and the crystalline directions are indicated.  
\begin{figure}
\centerline{\psfig{figure=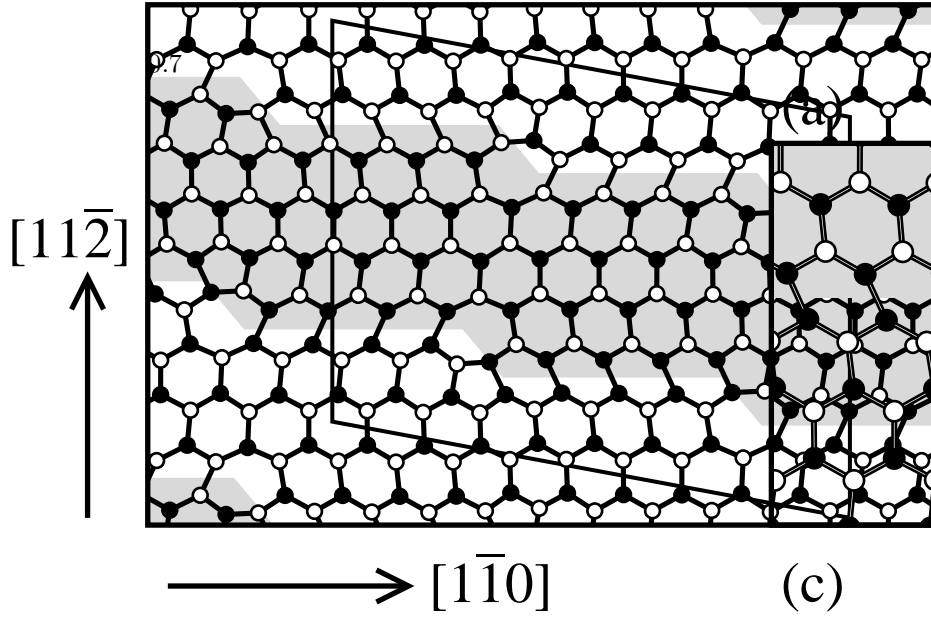,scale=0.7}}
\caption{
Supercell used for kinks and kink-soliton complexes. Crystalline
directions and supercell vectors indicated. Unit cell contains 864
atoms.}
\end{figure}
In
Figs.~3(a)-(e) the local structures of the five different types of
kinks are displayed. The notation we chose to name each kink type is
related to the orientation of the reconstructed bond, as one moves
from left to right in each diagram in Fig.~3. For example, in
Fig.~3(a) we denote the orientation at the left of the kink as
``right'' (R). Hence, we call this a right-left (RL) kink, the
notation following accordingly for the other types.  The above
supercell was used to compute the energies for the LR and RL
kinks. The lattice vector was staggered by twice the ``kink
vector''\cite{k-vec}, as shown in Fig.~2, in order to accommodate one
RL and one LR kink in each dislocation. Fig.~3(f) shows a complex of a
LR kink and a soliton. For the configurations in Figs.~3(c)-(f) the
bonds on either side of the defect have the same orientation. For
these, a supercell of 432 atoms was used (half of that shown in
Fig.~2), having one kink or complex in each dislocation, with a
supercell vector staggered by one kink vector.
\begin{figure}
\centerline{\psfig{figure=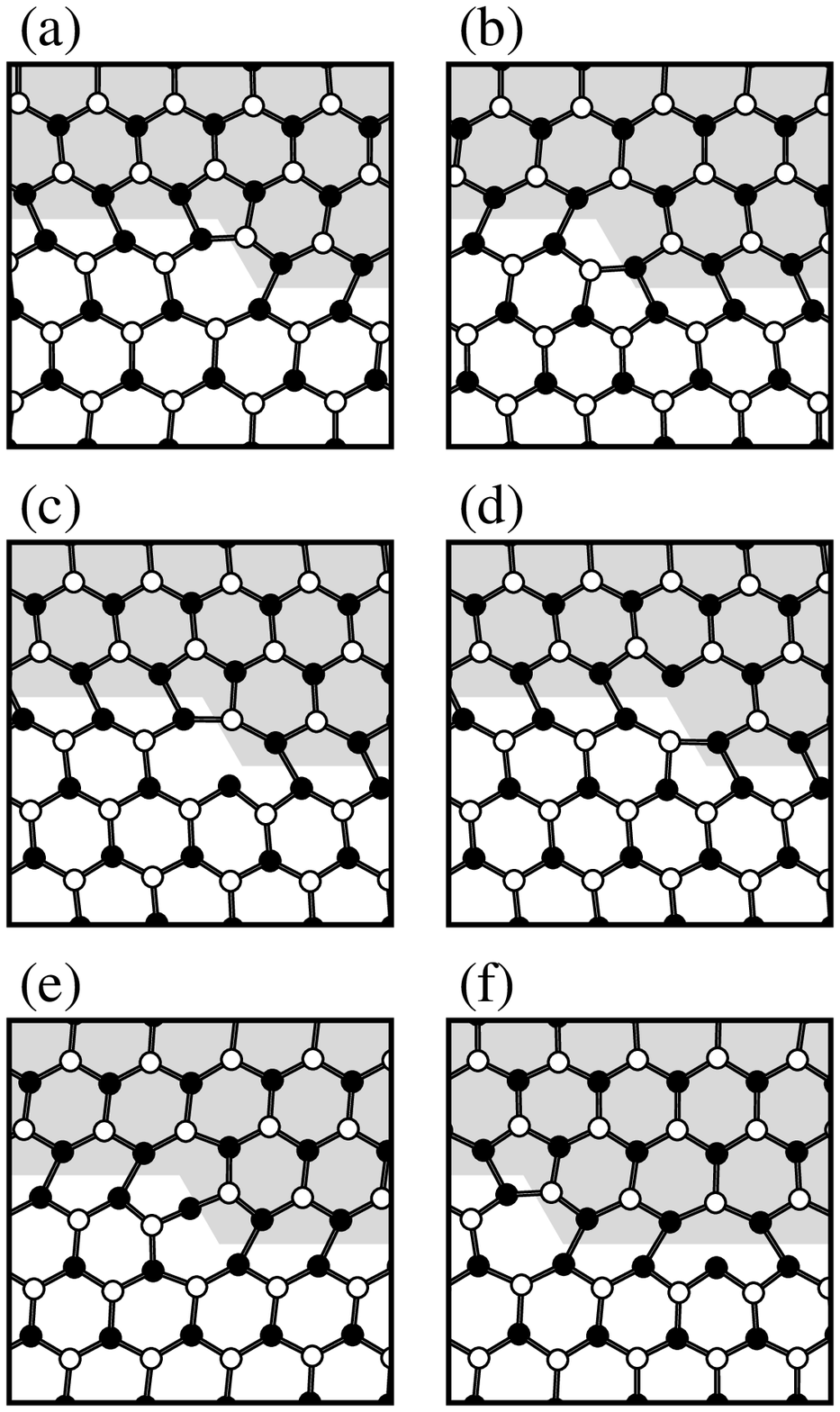,scale=0.7}}
\caption{
Core structure of various kinks and a kink-soliton complex. Kink
notation relates to bond reconstruction on each side of the defect,
and is explained in the text.  (a) RL kink.  (b) LR kink.  (c) LL
kink.  (d) LL$^*$ kink.  (e) RR kink.  (f) LR kink + soliton.}
\end{figure}

Our results for the energies of the configurations in Fig.~3 are shown
in Table~I. It is seen that the LR and RL kinks are much lower in
energy than all the others. (The LL and LL$^*$ kinks are found to be
unstable against emission of a soliton, a mechanism that is discussed
in detail in the next paragraph.) An inspection of Fig.~3 shows that
the LR and RL kinks are fully reconstructed, with no dangling
bonds. On the other hand, the high-energy RR kink and the unstable LL
kinks all contain a dangling bond (note the three-fold coordinated
atom at the core of each defect). This distinction is clearly
responsible for most of the energy difference.  On the basis of the
energetics alone, we can thus conclude that any kinks which occur in
the 90$^{\circ}$ partial dislocation in silicon will be almost
entirely of type LR or RL. This means that the orientation of the core
reconstruction is predicted to alternate from one inter-kink segment
to the next along the dislocation.

Moreover, we see that the RL + S and LR + S complexes (S = soliton)
have lower energies than the RR and LL kinks. From the energetics in
Table~I we can write the following reaction equations:
\begin{eqnarray}
&& RR \rightarrow RL + S + 0.41\;eV; \\
&& RR \rightarrow LR + S + 0.36\;eV; \\
&& LL \rightarrow LR + S\;\;\hbox{\it(spontaneous)};\\
&& LL^* \rightarrow RL + S\;\;\hbox{\it(spontaneous)}.
\end{eqnarray}
Reactions (1) and (2) above consists of the emission of a soliton by
the RR kink, which turns into a LR or RL in the process. We estimated
the energy barrier for these processes, obtaining the value of 0.05
eV. This result is not surprising in view of the low energy barrier
for soliton motion. The same mechanism is involved in the instability
of the LL and LL$^*$ kinks. Here the asymmetry of the local strain
fields is enough to remove the barrier to emission of the soliton in
one direction.

We decided to put the above picture to test by performing a MD
simulation on a 864-atom supercell containing four RR kinks, two in
each dislocation. Thus it should be possible for the system to convert
these kinks into alternating LR and RL kinks, as is to be expected
from the energetics in Table~I. At room temperature a RR kink in one
of the dislocations emitted a soliton after 0.2 ps, turning into a RL
kink. This soliton propagated towards the other RR kink in the same
dislocation and fused with it (converting it to a LR kink) after only
0.7 ps.  The latter process is equivalent to recombination of the
propagating soliton with an anti-soliton ``embedded'' in the RR kink.
\begin{table}
\caption{Calculated formation energy for defects in the core of the
90$^{\circ}$ partial dislocation in silicon. Included also are the
migration barriers for a soliton and the low energy kinks.}
\begin{tabular}{lcc}
                             &Formation energy
                                          &Migration barrier \\
                             &(eV)        &(eV)             \\
\tableline
soliton                      &\dec1.31    &\dec0.04    \\
LR kink                      &\dec0.50    &\dec1.87    \\
RL kink                      &\dec0.50    &\dec1.83    \\
LL kink\tablenote{Approximate energy. Defect is unstable}
                             &\dec1.74    &$\;\;$---\\
LL$^*$ kink\tablenotemark[1]
                             &\dec1.76    &$\;\;$---   \\
RR kink                      &\dec2.04    &$\;\;$---   \\
soliton + LR kink            &\dec1.68    &$\;\;$---   \\
soliton + RL kink            &\dec1.63    &$\;\;$---   \\
\end{tabular}
\end{table}

In Table~I, we also include the energy barriers to motion of the RL
and LR kinks. For metals, the formation of double kinks controls the
rate of dislocation motion, and the energy barriers to kink
propagation along the dislocation line are very small. The high values
we obtained for silicon are a signature of the highly directional
bonds in covalent semiconductors. In these systems, dislocation motion
is believed to be controlled by the kink mobility. Recent
experiments\cite{gotts,farber} have confirmed this picture, but some
controversy still remains, and the role of impurities as obstacles to
kink motion is yet to be fully explored. Here we concentrate on the
important issue of the size of the kink formation energies and
migration barriers, and make a connection with the experimental
results. For the velocity of a gliding dislocation we have, from the
theory of Ref.\cite{h&l}
\begin{equation}
v_d \propto exp[-(U_k + W_m)/kT]\;,
\end{equation}
where $U_k$ is the formation energy of a kink and $W_m$ is the energy
barrier for kink migration along the dislocation. The experimental
estimates based on transmission electron microscopy (TEM) or
intermittent loading (IL) measurements\cite{gotts,farber} range from
0.40 to 0.62 eV for $U_k$ and 1.50 to 1.80 eV for $W_m$. Our results
from Table~I, $U_k = 0.50$ eV and $W_m = 1.85$ eV (average between LR
and RL values), fall within the range of the experimental numbers.
The theory in Ref.\cite{h&l} can be used to
calculate the Peierls stress of materials. When the above numbers are
used for silicon, one obtains a value that is too low in comparison
with results of high-stress measurements. Discrepancies are also found
for quantities such as the velocity of steady-state motion of a
dislocation under static load\cite{farber}. It has been argued in
Ref.\cite{farber} that these discrepancies are to be assigned to point
defects that would influence both the concentration of kinks and their
migration barriers, by creating inhomogeneities in the potential
relief (the potential felt by a moving dislocation due to the
periodicity of the lattice). The above comparison between experiment
and our results, which are valid for a homogeneous potential relief,
reinforces the plausibility of this scenario.

In summary, we investigated different structural and dynamical
properties of the 90$^{\circ}$ partial dislocation in silicon. We
verified that the core undergoes a reconstruction that breaks the
mirror symmetry along the dislocation line, as previously
reported. This leads to the existence of solitons which are shown to
be highly mobile along the dislocation core, and to a multiplicity of
kinks whose stability is found to depend, in each case, on the
reconstruction of dangling bonds at the core of the defect. We find
that the high-energy kinks transform into low-energy ones by soliton
emission. The energy barriers associated with this mechanism are small
or absent in each case. The low-energy kinks are fully reconstructed
with no dangling bonds, and impose an alternation of the orientation
of the core reconstruction from one inter-kink segment to the
next. The mobility of these low-energy kinks presumably determines the
dislocation mobility in the glide plane. Our calculated formation
energies and energy barriers for these kinks are in good agreement
with available experimental estimates.

During the final stages of preparation of this manuscript, we became
aware of related work by Hansen {\it et al.} \cite{hansennew}, who
consider some similar kink structures but with a much smaller
interkink separation than was considered here.

This work was supported by NSF Grant DMR-91-15342.  R.~W.~Nunes
acknowledges the support from CNPq - Brazil.  J.~Bennetto acknowledges
support of ONR Grant N00014-93-I-1097.


%

\begin{references}
\bibitem{exact} Because the functional is exactly a cubic polynomial,
its minimum can be located immediately by evaluating its value and
first three derivatives at any single point on the line.
\bibitem{jonesexplan} This is in reasonable agreement with the
estimate of 1.2 eV obtained in Ref.\cite{jones} from a
Si$_{34}$H$_{39}$ cluster calculation.
\bibitem{k-vec} A ``kink vector'' is the vector by which a dislocation
shifts between two neighboring Peierls valleys. It relates equivalent
atoms in adjacent sub-layers of the diamond lattice.
\end{references}
\end{document}